\documentclass[useAMS,usenatbib]{mn2e}
\usepackage{graphicx}

\title[Star formation triggering near groups and clusters]{Triggering of merger-induced starbursts by the tidal field of galaxy groups and clusters}
\author[M. Martig and F. Bournaud]{M. Martig$^{1}$\thanks{E-mail: marie.martig@cea.fr} and F. Bournaud$^{1}$\\
$^{1}$Laboratoire AIM, CEA/DSM - CNRS - Universit\'e Paris Diderot. Dapnia/SAp, 91191 Gif-sur Yvette, France}
\begin{document}

\date{Accepted xxx. Received xxx; in original form xxx}

\maketitle

\newcommand{\normal}{}

\begin{abstract}
Star formation in galaxies is for a part driven by galaxy mergers.
At low redshift, star formation activity is low in high-density
environments like groups and clusters, and the star formation
activity of galaxies increases with their isolation. This star
formation -- density relation is observed to be reversed at $z \sim
1$, which is not explained by theoretical models so far. We study
the influence of the tidal field of a galaxy group or cluster on the
star formation activity of merging galaxies, using N-body
simulations including gas dynamics and star formation. We find that
the merger-driven star formation is significantly more active in the
vicinity of such cosmological structures compared to mergers in
the field. The large-scale tidal field can thus enhance the activity of
galaxies in dense cosmic structures, and should be particularly
efficient at high redshift before quenching processes take
effect in densest regions.
\end{abstract}


\section{Introduction}

In the Local Universe ($z \simeq 0$), the strongest starbursts occur
in interacting systems; luminous and ultraluminous infrared
galaxies are usually found in major mergers involving at
least two galaxies of comparable masses \citep[e.g.,][]{D97}. At redshift
$z\simeq 1$ and above, an important fraction of starbursts still
seem associated with interactions and mergers
\citep{conselice03,bridge07}, even if the bulk of star
formation may not be simply merger-driven \citep{bell05, daddi07,
jogee07}. Theory and numerical simulations succeed in explaining the
triggering of starbursts by galaxy interactions and mergers
\citep[e.g.,][]{MH96}. However, supernovae feedback can regulate the star
formation in mergers \citep{cox06} and \citet{dM07} showed from a
statistical study of a large sample of galaxy interactions and mergers
 that the maximal star formation rate (SFR) in interacting galaxies is rarely higher
than a few times that of isolated galaxies even in equal-mass
mergers -- the activity decreases significantly with increasing mass
ratios (Cox et al. 2007). This suggests that some factors
contributing to the triggering of high SFRs in mergers could have
been neglected.

At low redshift, the star formation activity in a given galaxy is
anti-correlated to the density of galaxies that surround it
\citep{lewis02,kauffmann04}, galaxies in or near groups forming less
stars than galaxies in poorer regions of the field. In clusters,
star formation is even less active, which is explained by a variety
of phenomena including the ram-pressure stripping \citep{quilis00},
galaxy harassment \citep{moore96}, and galaxy strangulation
\citep{kawata07}.

Cosmological $\Lambda$CDM models \citep[Millenium,][]{springel05}
explain the Local star formation activity -- environmental density
relation, but predict that it gets reversed only at very high
redshift $z>2$. Yet, it has been recently discovered that the star
formation -- density relation is already reversed at $z \simeq 1$,
where the star formation activity of galaxies increases with the
local density of the surrounding galaxies, except in the very
densest regions (Elbaz et al. 2007, Cooper et al. 2007). This
reversal of the star formation-density relation at $z\sim 1$ is
theoretically unexplained by hierarchical models and cannot simply
result from major mergers being more frequent in dense environments
(see Elbaz et al. 2007). This suggests that unknown environmental
mechanisms can trigger the star formation activity further than what
mergers do. These environmental processes may still take place
during mergers, at least for a part, since the later remain a major
driver of star formation.

In this Letter, we study the influence of the tidal field of 
galaxy groups and clusters on the star formation activity of
isolated and merging galaxies. While the tidal field of such
structures alone only triggers a weak activity in a single
galaxy, we show that it can strongly enhance the SFR of
merger-induced starbursts. Mihos (2004) has shown that an external
potential can modify the morphology of gaseous tidal
tails developed in galaxy mergers, but did not study the star
formation in this context. Here we show that galaxy mergers are
statistically more efficient in triggering strong starbursts if they
take place in the vicinity of a larger structure. This should
contribute to the triggering of star formation in dense environments
at $z \sim 1$. Indeed the triggering of star formation by the
large-scale tidal field should be efficient in particular in young
groups and near forming clusters, where the quenching factors did
not have time to act yet. The numerical simulations are described in Section~2, and the
results are presented in Section~3. Our conclusions are discussed and summarized in
Sections~4 and~5.

\section{Numerical simulations}

\subsection{Code and model description}
Galaxies are modelled as stars, gas and dark matter particles. The
gravitational potential is computed with an FFT-based particle-mesh
technique, with a spatial resolution and softening of 190~pc, as described in
 \citet{BC02, BC03}. Gas dynamics is modelled with a
sticky-particles scheme with parameters $\beta_r$=0.7 and $\beta_t$=0.5. Star formation is computed using a local Schmidt law : the star
formation rate is proportional to the local gas density to the exponent 1.5 \citep{kennicutt98}.

In order to directly compare the SFR of an interacting galaxy and of the same 
galaxy isolated, at fixed gas mass, we implement a gas disc of $10^5$ particles 
in this galaxy, together with 7$\times10^4$ stellar particles and 5$\times10^4$ dark matter particles. The particle mass resolution is 1.5$\times10^4$~M$_{\sun}$ for gas, 1.8$\times10^5$~M$_{\sun}$ for stars, 10.8$\times10^5$~M$_{\sun}$ for dark matter.
The merging companion is modelled with 7$\times10^4$ stellar particles and 5$\times10^4$ dark matter particles.
Because we study star formation mainly in
a context of $z>0$, we model galaxies with a moderate visible
 mass of $1.5 \times 10^{10}$~M$_{\sun}$. At $z=0$ this 
corresponds to somewhat small spirals (similar to M~33).
The bulge:disc mass ratio is 0.24, the gas mass fraction
in the disc is 15\%. The bulge has a scale-length of 600~pc, the
disc has a Toomre profile with a radial scale-length of 1.6~kpc for
stars and 4.6~kpc for gas, truncated at 5.6~kpc. A dark halo of mass
$5.4 \times 10^{10}$~M$_{\sun}$ is implemented with a Plummer
profile of scale-length 6~kpc truncated at 20~kpc, giving a circular 
velocity $V_{\mathrm{circ}} \simeq 100$~km~s$^{-1}$.

Galaxies are evolved as isolated systems for 500~Myr before
simulations are started. This way, galaxies already have acquired a realistic (barred) spiral structure when
they interact and merge, without having time to undergo a major secular evolution of their bulge mass or disc size.
Star formation is shut down during this initial period, so that interactions really start with the gas fraction indicated above.
The evolution of the SFR is thus related to the interaction/merger, without any bias introduced by the transition from the initial axisymmetric model to a realistic spiral disc.

\subsection{Galaxy mergers, group and cluster potential}

We performed simulations of binary equal-mass mergers of two spiral
galaxies corresponding to the model above. The orbital parameters of
the merging pair are as follows:

\begin{itemize}
\item the inclination $i$ of the orbital plane with respect to each galactic disc was fixed to 33 degrees. This is the average value $\int i\; p(i)\; di$, the probability of an inclination $i$ being $p(i)\propto i$ in spherical geometry. This way we model typical orbits that are not coplanar nor polar.
\item the initial velocity $V$ was varied to 0.2, 0.4, 0.6 and 0.8, in units of the circular velocity $V_{\mathrm{circ}}$.
\item the impact parameter $b$ (defined as the initial separation in the direction perpendicular to the initial velocity) was varied to 3, 4, 5, and 6 times the gas disc radius.
\item the orientation was varied to prograde and retrograde.
\end{itemize}

We model groups and clusters gravitational potential using a
Plummer profile. This choice is discussed in Section 3.3, and should
be representative of most groups and clusters tidal field at least in the
peripheral regions studied here. The modelled cluster has a mass of
$10^{15}$~M$_{\sun}$, and a radial scale-length of 400~kpc (this
choice would be reasonably representative for instance of the Virgo
cluster \citep[e.g.,][]{fouque01} and the group a mass of
$5\times10^{13}$~M$_{\sun}$ and a scale-length of 150~kpc, which
could be representative of the Local Group depending on its
dark:visible ratio.

The galaxy pair was initially placed at 400~kpc from the centre of
the cluster (resp. 150~kpc from the group centre). Galaxies are not
placed specifically in central regions, but in the periphery. This
is a more general choice, and in the case of clusters it ensures
that galaxies there can still contain gas reservoirs and form stars.

We chose four possible configurations for the relative position of
the galaxy pair and the group or cluster:
\begin{itemize}
\item configuration 1: the group/cluster centre is in the orbital plane, along the axis supporting the initial relative velocity of the galaxy pair.
\item configuration 2: the group/cluster centre is in the orbital plane, along the axis perpendicular to the initial relative velocity of the galaxy pair.
\item configuration 3: the group/group centre is in the orbital plane, along the bisector of the two previous directions.
\item configuration 4: the group/cluster centre is at 45~degrees from the orbital plane, with a projected position in the orbital plane similar to configuration 3.
\end{itemize}

Each orbital parameter for the merging galaxies has been simulated without any external field, and with the group and the cluster in each configuration; the total number of cases is then as large as 320. We restrict ourselves to cases leading to a merger, otherwise the parameter space to explore would be too large. We make the choice of the galaxy pair having no initial velocity w.r.t. the group/cluster (but free to move within in) as justified in Sect.~\ref{312}.

\section{Results}

In the following, `relative SFR' refers to the SFR of a galaxy
with a merging companion and/or a group or cluster tidal field,
divided by the SFR of the same galaxy isolated. The starbursts are described with the maximum value of the relative SFR, i.e. SFR$_{\mathrm{peak}}$/SFR$_{\mathrm{isol}}$ on Fig.~\ref{fig_pair}.

\subsection{Single galaxy in a group/cluster tidal field}

\begin{figure}
\centering
\includegraphics[width=8cm]{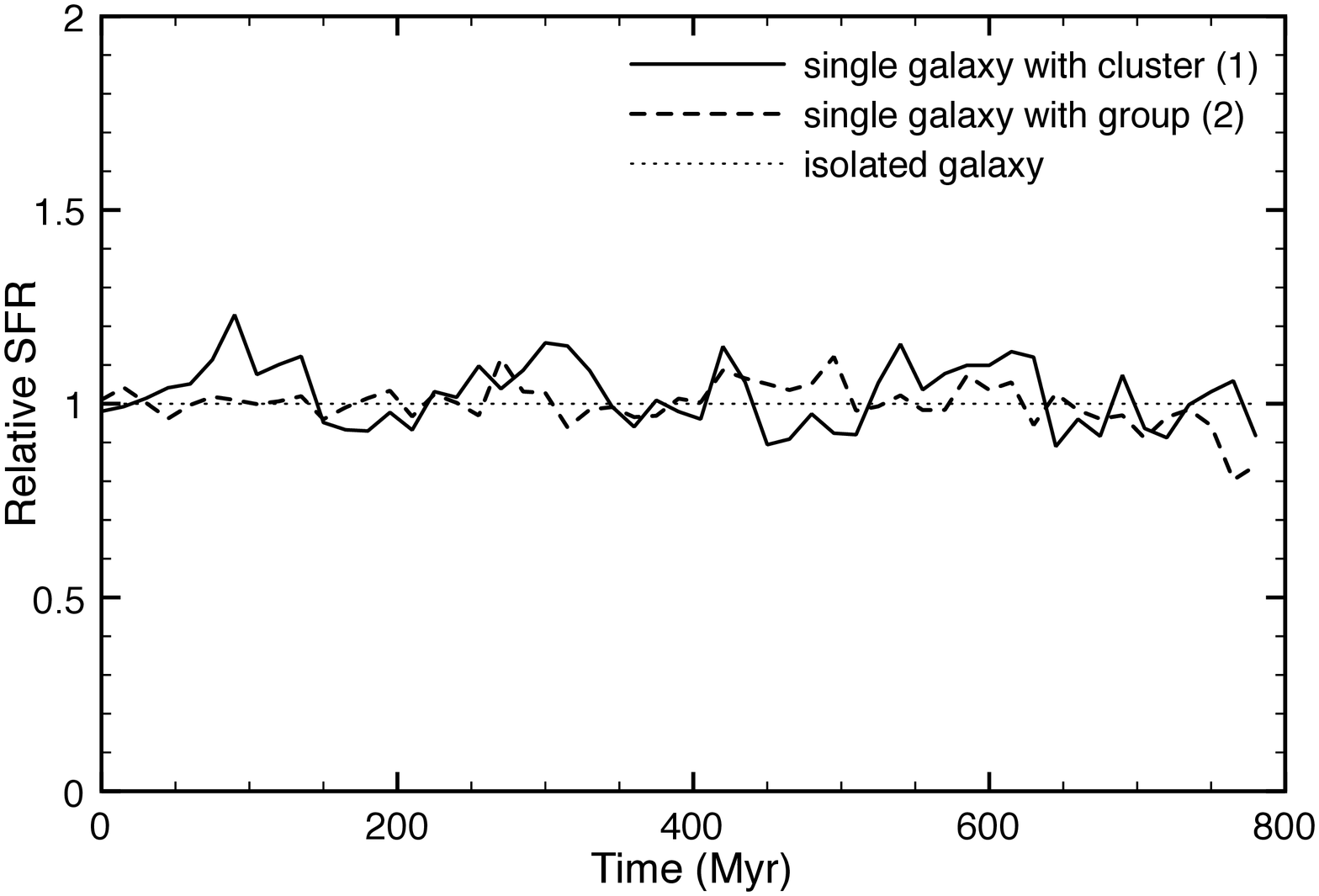}
\caption{SFR of a single galaxy with a cluster in configuration 1 or a group in
configuration 2, relative to the SFR of the same galaxy without the
external field.}\label{fig_isol}
\end{figure}

We show in Fig.~\ref{fig_isol} the evolution of the relative SFR for
a single galaxy in the group and cluster tidal fields (without any
interacting companion) for two of the simulated configurations. The average
relative SFR is in both cases only slightly larger than
1. Thus, the tidal field of the cluster or
the group can weakly trigger the star formation in a
single galaxy, but without driving significant starbursts as mergers could do.

\subsection{Merging pair in a group/cluster tidal field}\label{312}

\begin{figure}
\centering
\includegraphics[width=8cm]{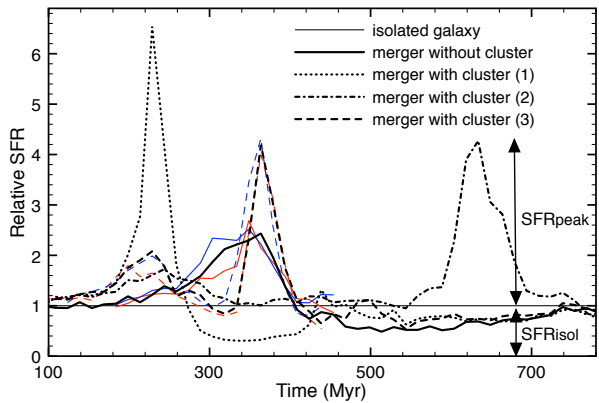}
\caption{Comparison of the SFR (relative to the isolated case) for mergers without and with a cluster, in different configurations. Parameters are described in text (Sect.~\ref{312}). The thin blue curves show SFR evolutions in the target galaxy when the perturbing galaxy also contains gas; the red curves show test simulations with a mass resolution of dark matter particles enhanced by a factor 3.}\label{fig_pair}
\end{figure}

The effect of the group or cluster tidal field on star formation can
be larger when we replace the single galaxy by a merging pair.
Before the systematical statistical study of Sect.~\ref{32}, we show
here an example of galaxy pair in the cluster field.
Fig.~\ref{fig_pair} shows the relative SFR of a galaxy merging with
an equal mass companion on a direct orbit with $V=0.6V_{\mathrm{circ}}$ and
$b=4R_{\mathrm{disc}}$, with and without the cluster tidal field. The galaxy merger induces a starburst, the intensity of which is
significantly affected by the cluster field: the peak intensity of
the relative SFR is increased by the presence of the cluster, at
various levels depending on the configuration. The triggering can be
quite significant, with for example on Fig. \ref{fig_pair} a peak
SFR increased by a factor three when the cluster is in configuration
1, compared to the merging pair without the cluster field.

We tested in such simulations the influence of the initial velocity
of the galaxy pair w.r.t the group/cluster, starting with radial and
tangential velocities of 100~km.s$^{-1}$ and 300~km.s$^{-1}$.
The change in the SFR is found to be within five per cent.
This weak influence justifies our choice to not vary systematically
the initial velocity of the merging galaxy pair w.r.t the
group/cluster.

\subsection{Statistical analysis: star formation triggering}\label{32}

To perform a statistical analysis on our whole simulation sample,
we take into account the fact that the various configurations leading to mergers
that we simulated have different likelihoods, and must then be
weighted accordingly. Within the simple assumption of a random distribution of companions, the collision rate varies as the velocity and the cross-section $\pi b^2$ for an impact parameter $b$. Thus, each simulation should be attributed a probability $\propto b^2 V f(V)$
(e.g., Mihos 2004) where $f(V)$ is the velocity distribution
of galaxies. The distribution $f$ is generally unknown. The real
distribution $f(V)$ should increase with $V$ (in particular in
clusters, and for the moderate velocities leading to mergers that we
study). In the following we present the results for $f(V)=1$ and
$f(V)\propto V$, assuming that the real distribution is likely in
between. The variations caused by different assumptions on $f(V)$
are anyway small (see values below), and we mainly focus on the
$f(V)=1$ assumption that is found to give a conservative limit to the final
result, the conclusions being quantitatively stronger (in minor
proportions) if we assume an increasing form for $f$.

\begin{figure}
\centering
\includegraphics[width=8cm]{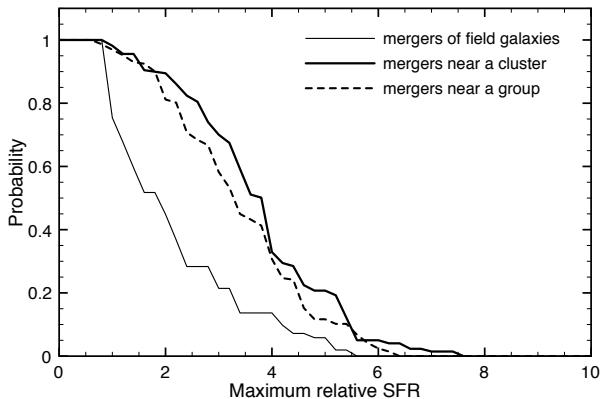}
\caption{Probability to have a maximum relative SFR larger than the
value specified on the x-axis for mergers of field galaxies vs
galaxies in a group or cluster tidal field. The results shown are
for the conservative assumption $f(V)=1$.}\label{fig_stat}
\end{figure}

We show on Fig.~\ref{fig_stat} the statistical distribution of the
maximum relative SFR (SFR$_{\mathrm{peak}}$/SFR$_{\mathrm{isol}}$ on Fig.~\ref{fig_pair}) for merging galaxy pairs with/without the external
gravitational potential. We notice that the major mergers are
significantly more efficient to trigger star formation if they take
place in the gravitational field of a cluster or a group. Indeed,
for a merging pair in the field, the fraction of `significant'
bursts (SFR multiplied by a factor 2 at least compared to the
isolated reference disc) is 45\% for $f(V)=1$ (respectively 35\% for
$f(V)=V$). This fraction becomes 81\% (resp. 85\%) for the model
group potential and  90\% (resp. 92\%) in the periphery of the model
cluster. Similarly for `strong' starbursts (say, SFR multiplied by
at least a factor 5), only 6\% (3\%) of the mergers without any
external potential reach this level, while this fraction is increased
to 11\% (21\%) in a group tidal field as well as in a cluster tidal
field. Thus, starbursts at all levels are triggered by the presence
of the group/cluster potential: over our sample of mergers,
the maximal instantaneous SFR is on average doubled, depending on
the shape of $f(V)$, compared to mergers in the field far from
groups and clusters. The enhancement is even larger in some cases,
in particular for the highest initial velocities. This triggering effect is also important if we consider the
integrated star formation over the burst duration (i.e. the total
mass of stars formed) that is on average multiplied by 1.3--1.4 in
the vicinity of both the model group and cluster.

We thus find a noticeable enhancement of the star formation activity
resulting from the addition of the cluster or group potential. This
effect is comparable in amplitude to the primary effect of star
formation triggering by a major merger compared to an isolated
galaxy, hence a significant one. Nevertheless the quantitative
results are exact only for the adopted group or cluster mass,
radius, and distance of the merging pair to the group/cluster
centre. Other structures would be more or less efficient in
triggering star formation depending on the intensity of their tidal
field on the merging galaxies.

\section{Discussion}

\subsection{Modelling and numerical issues}

We have assumed a Plummer profile for the potential of the modelled
group and cluster. In the case of clusters, real potentials are more
likely cuspy \citep{pointecouteau05}. A cuspy Hernquist profile with
the same half-mass radius and total mass as our Plummer profile
would actually exert a larger tidal field in its central regions near the cusp, 
and a comparable one (but always slightly larger) in the outer regions. 
The Plummer profile is thus a conservative choice, which could 
only lead to somewhat underestimating the effect. Changing the 
mass or scale-length of the structure by a factor two would lead to 
similar variations in its tidal field as changing its profile.
Our choice was also motivated by the fact that forming groups and
clusters at $z \sim 1$ are likely made-up of several merging
sub-groups, with likely a large-scale profile flatter than relaxed
structures. The halo profile of each galaxy might also be different from the one
assumed in our model. This however should not have a major influence
because the dark matter does not dominate the mass distribution in
the central regions where the gas flows and star formation are
triggered.

We chose to model the perturbing galaxy as gas-free, because merger-induced starbursts are mostly driven by gravity torques rather than hydrodynamical processes \citep[e.g.,][]{MH96}. Some test simulations (see examples on Fig.~\ref{fig_pair}) confirm that the SFR evolution of the studied target galaxy is little influenced by the gas content of the perturbing galaxy even when the gas fraction in the latter is as high as in the former.

A sticky-particles code was employed to model the dynamics of the
interstellar gas. The purpose is to provide a kinematically cold
medium when the galaxy interaction occurs, but it is the response of
the interstellar medium to the galaxy interaction that drives a gas
inflow and subsequent starburst -- not the sticky-particles
dynamics. It takes several billion years for the angular momentum
and energy of the ISM to vary significantly \citep[see appendix in][]{BC02} and so the gas inflows and bursts of star formation are
not largely dependent on the choice of this model and related
parameters. We also note that our results on starbursts in merging pairs outside external potentials compare
favourably to those obtained by \citet{dM07} with an SPH model, and to the moderate SFR enhancement in mergers compared to
non-interacting galaxies in GEMS \citep{jogee07}.
Furthermore, a limited mass resolution of dark matter particles can induce N-body scattering, which can affect the structure of numerical merger remnants \citep[e.g.,][]{naab99}. We performed tests with a number of dark matter particles increased by factor three, which is found to leave the SFR evolution unaffected (Fig.~\ref{fig_pair}).

\subsection{Conditions for a star formation enhancement}

Our results show that a given galaxy pair that merges in the
vicinity of a group or cluster has an SFR enhanced by this external
tidal field. In the statistical analysis, we have compared results
assuming a similar velocity distribution for galaxies in the field
and those near groups/clusters (either $f(V)=1$ or $f(V)=V$ in all
cases). This should be close to reality in groups, but one could
expect higher velocities to dominate in the vicinity of clusters.
Taking this into account would not however result in major changes; the 
average SFR enhancement is affected by 10\% if we assume $f(V)=1$ in the
field and $f(V)=V$ near the cluster. One also expects an increase in
the proportion of fly-bys not followed by mergers in the vicinity of
clusters (not necessarily near groups). \cite{dM07} showed that
the SFRs are of the same order of magnitude for mergers and fly-bys, so that
our results should be at first order extendable to higher-velocity fly-bys. When weighting our results with the cross-section $\pi b^2$, the impact parameter $b$ was simply estimated at the beginning of our simulations ($t=0$). The external tidal field would actually modify the orbits even before $t=0$ and possibly bias this parameter $b$ compared to the real impact parameter $b_{\infty}$ at infinite distance. However, the difference between $b$ and $b_{\infty}$ is minimized when the group/cluster is in configuration 1 (along the direction of the initial velocity), and a significant enhancement of the SFR is still found when we restrict ourselves to this configuration (even 20\% higher than in the other configurations). The star formation triggering that we found cannot then be an artefact resulting from the assimilation of $b$ to $b_{\infty}$.

The comparable effect found in our model group and model cluster
indicates that the main requirement is the presence of a tidal
field, while the total size and mass of the structure play only a
secondary role in determining the exact level of triggering. One can
however wonder if this really implies an observable triggering of
the SF activity, because the merging pairs in dense and low-density
regions are not necessarily similar. Galaxies in groups can still
contain important gas reservoirs, but galaxies in the central
regions of relaxed clusters at $z \sim 0$ are mostly gas-depleted.
The triggering effect of the large-scale tidal field should thus
mainly affect galaxies in moderate density environments like groups,
galaxies in the periphery of relaxed clusters but not at their
centre, or in young/forming clusters where the quenching mechanisms
did not have time to act yet. Galaxies in the periphery of clusters
can indeed still have large gas reservoirs (Egami et al. 2006, see
also Chung et al. 2007 for Virgo spirals). In the light of our
results, we may then speculate that the specific SFR of galaxies
(at least interacting ones) in the outer regions of clusters is higher than
in both the field and the central cluster regions.
Statistical comparison of the outskirts of clusters compared to the
field and to the central cluster regions could confirm this prediction.

\section{Conclusion}

In this paper, we have shown that a major galaxy merger is more
efficient to trigger an intense burst of star formation if it takes
place in the tidal field of a galaxy group or cluster. While the
group/cluster fields do not trigger much the star formation in a
single galaxy, the effect on merging pairs is important. The
starbursts in our simulations are amplified by a factor 2 on average
(sometimes much more) for typical groups and clusters. More massive
structures could even have larger quantitative effects depending on
the intensity of the tidal field. A pair of M33-like spiral galaxies
merging in the vicinity of a Local-like group or a Virgo-like
cluster would see the intensity of its starburst amplified by the
external field by typically a factor $\sim 2$, but possibly more on
some orbits or regions of high tidal field. In a forthcoming paper, we will study the dynamical
response of gas during mergers within such an external tidal field and the 
connection with the enhancement of starbursts.

Dense cosmological structures trigger the merger-induced star
formation by the action of their tidal field. This holds at least
from groups to clusters, the level depending on their mass and
size. Because gas dynamics and star formation must be spatially well
resolved for SFRs to be accurately computed in galaxy mergers
, large-volume cosmological simulations may miss
or underestimate this effect. This triggering of merger-induced
starbursts by the tidal field of dense cosmological structures
should be particularly efficient at high redshift ($z \simeq 1$ and
above), when for instance the clusters begin to form and exert a
tidal field, but are not virialized yet. Later on at $z \simeq 0$,
quenching mechanisms have acted in the highest density regions resulting in a less active star formation there.

Our results unveil a new star-formation triggering mechanism in groups
and at the periphery of clusters, which can act in particular at
high redshift before star formation is quenched in dense regions.
This can contribute to explain why LIRGs are often found in
proto-cluster environments \citep{laag06} and the high frequency of
blue star forming galaxies in young high-redshift clusters
\citep{butcher-oemler}. More generally, this can trigger the
star formation activity in group mergers, and contribute to the reversal of the star formation -- density relation with increasing
redshift.

\section*{Acknowledgments}
Simulations were performed on NEC-SX8 vector computers at CEA/CCRT and CNRS/IDRIS, as part of the Horizon project. Discussions with David Elbaz, Pierre-Alain Duc, Giovanna Temporin and Emanuele Daddi motivated this study and improved
the manuscript. We are grateful to Yves Revaz for providing 
his visualisation software pNbody, Paola Di~Matteo, Shardha Jogee, Chanda Jog, Fran\c{c}oise Combes and Romain Teyssier for useful discussions, and to an anymous referee for constructive remarks.

\end{document}